\documentclass[
  manuscript=letter,
  year = 2026,
  volume = 1,
  journal= 
]{cup-journal2}

\usepackage{amsmath}
\usepackage[nopatch]{microtype}
\usepackage{booktabs}
\usepackage{url}
\usepackage{float}
\usepackage{soul}
\usepackage{pdfpages}

\title{Correcting for Nonignorable Nonresponse Bias in Ordinal Observational Survey Data}

\author{Lukáš Lafférs}
\affiliation{Department of Mathematics, Matej Bel University, Banská Bystrica, Slovakia}
\alsoaffiliation{Department of Economics, Norwegian School of Economics, Bergen, Norway}

\author{Jozef Michal Mintal}
\affiliation{Research and Innovation Center, Matej Bel University, Banská Bystrica, Slovakia}
\email[F. Author]{jozef.mintal@umb.sk}

\author{Ivan Sutóris}
\affiliation{National Bank of Slovakia, Bratislava, Slovakia}


\begin{document}

\begin{abstract}
Many political surveys rely on post-stratification, raking, or related weighting adjustments to align respondents with the target population. But when respondents differ from nonrespondents on the outcome itself (nonignorable nonresponse), these adjustments can fail, introducing bias even into basic descriptives. We provide a practical method that corrects for nonignorable nonresponse by leveraging response-propensity proxies (e.g., a respondent's rating of the interview, or interviewer-coded cooperativeness) observed among respondents to extrapolate toward nonrespondents, while directly integrating observable covariates and retaining the benefits of post-stratification with known population shares. The method generalizes the variable-response-propensity (VRP) framework of Peress (2010) from binary to ordinal outcomes, which are widely used to measure trust, satisfaction, and policy attitudes. The resulting estimator is computed by maximum likelihood and implemented in a compact R routine that handles both ordinal and binary outcomes. Using the 2024 American National Election Study (ANES), we show that accounting for nonignorable nonresponse produces substantively meaningful shifts for life satisfaction (estimated latent correlation $\rho \approx 0.47$), while yielding only modest changes for retrospective economic evaluations ($\rho \approx 0.14$), highlighting when nonignorable nonresponse substantively affects survey estimates.
\end{abstract}

\section{Motivation}
Survey data are foundational to empirical research in political science, yet rising nonresponse and selection bias threaten the validity of the inferences drawn from them \citep{Meng2018,CAVARI_FREEDMAN_2023,Jackman_Spahn_2019}. Widely used adjustments such as post-stratification or raking can align respondents with population margins on observed demographics, but they can remain biased when respondents differ systematically from nonrespondents on the outcome itself \citep{bailey_polling_2024,Bailey_2025}. 

The core problem is nonignorable nonresponse: the likelihood of responding depends on the
unobserved outcome itself. While several frameworks address this issue\footnote{See Bailey (2024) for a detailed overview.}, the variable-response-propensity estimator (VRP) proposed by \textcite{peress2010correcting} remains distinctive. Building on ``continuum of resistance'' logic, VRP leverages a response-propensity proxy observed among respondents (e.g., interviewer-coded cooperativeness, a respondent's rating of the interview, or related paradata) to extrapolate from low-propensity respondents toward nonrespondents, while directly integrating observable covariates into the correction framework and retaining the benefits of post-stratification with known population shares.

However, the original VRP estimator is framed for binary outcomes, while many key survey measures in political science, such as trust, satisfaction, and economic evaluations, rely on ordinal scales. Collapsing ordinal outcomes to binary discards information and can change substantive conclusions, especially if selection operates differently across the response scale \citep{stromberg_collapsing_1996,lauderdale_decomposing_2018}.

This letter provides a practical method for correcting nonignorable nonresponse in ordinal survey data when a response-propensity proxy is available, generalizing VRP beyond binary outcomes. The estimator is computed via maximum likelihood and implemented in a compact \textsf{R} routine covering both ordinal and binary outcomes, extending the original binary-only C++ implementation of \textcite{peress2010correcting}. Our method complements recent experimental approaches to nonignorable nonresponse correction \citep{Bailey_2025}, as it can be applied to observational surveys without requiring randomized instruments. It further complements the semiparametric approach of \textcite{li_correcting_2026}, which leverages callback data (outcome variation across successive survey contact stages) to achieve identification without distributional assumptions for binary outcomes, while our method can be applied to ordinal outcomes in surveys without a multi-stage callback design.

An illustration of our approach using the 2024 American National Election Study shows that accounting for nonignorable nonresponse can produce meaningful shifts for some outcomes (e.g., life satisfaction) but small or negligible changes for others (e.g., views on the death penalty), underscoring that the value of nonresponse correction is outcome-specific and empirically testable.

\section{Problem Formulation}

We now generalize the VRP framework \citep{peress2010correcting} for ordinal outcomes. Let $y_n \in \{1,2,\dots,Y\}$ denote an ordinal outcome of interest for unit $n$ and let $r_n$ denote an ordinal response-propensity proxy recorded among respondents, taking values in $\{1,2,\dots,R\}$ and ordered so that smaller values indicate higher propensity to respond. We embed unit nonresponse as an additional category $R{+}1$ in the response equation: $r_n=R{+}1$ corresponds to unit nonresponse, in which case $y_n$ is missing. The key parameter is $\rho$, the correlation between latent errors in the outcome and response equations, which captures nonignorable selection on the (unobserved) outcome after conditioning on covariates.

Operationally, we observe $(y_n,r_n,x_n,z_n)$ only for respondents ($r_n\le R$). Unit nonresponse enters through the number of missing units $N_{\text{miss}}$ (known when the sampling-frame response rate is available, or treated as a sensitivity parameter in public releases) and through population shares of strata defined by $z$: we partition the population into $K$ strata indexed by $k$ with covariate profile $z_k$ and population share $p_k^{z}$ (with $\sum_{k=1}^K p_k^{z}=1$).

We model the joint outcome and response processes using correlated latent-variable ordered probit models:
 \medskip \begin{equation}\label{eq:model} \begin{array}{c@{\hskip 1cm}c} \textit{Outcome model} & \textit{Response model} \\[0.8em] \begin{array}{l} y_n \in \{1,2,3,\ldots,Y\} \\[0.5em] y_n^* = \alpha^T x_n + \epsilon_n \\[0.5em] y_n = \begin{cases} 1 & \text{if } y_n^* \le \lambda_1 \\ 2 & \text{if } y_n^* \in (\lambda_1,\lambda_2] \\ 3 & \text{if } y_n^* \in (\lambda_2,\lambda_3] \\ \vdots \\ Y & \text{if } y_n^* > \lambda_{Y-1} \end{cases} \end{array} & \begin{array}{l} r_n \in \{1,2,3,\ldots,R, R+1\} \\[0.5em] r_n^* = \beta^T z_n + \eta_n \\[0.5em] r_n = \begin{cases} 1 & \text{if } r_n^* \le \theta_1 \\ 2 & \text{if } r_n^* \in (\theta_1,\theta_2] \\ 3 & \text{if } r_n^* \in (\theta_2,\theta_3] \\ \vdots \\ R & \text{if } r_n^* \in (\theta_{R-1},\theta_R] \\ R+1 & \text{if } r_n^* > \theta_R \end{cases} \end{array} \\[2em] \multicolumn{2}{c}{ \displaystyle \text{corr}(\epsilon_n,\eta_n) = \rho } \end{array} \end{equation} Errors $(\epsilon_n,\eta_n)$ are assumed to be jointly normally distributed with unit variances\footnote{The joint normal specification is related to bivariate latent-variable selection models studied more broadly. \textcite{marchenko_heckman_2012} replace bivariate normality with a Student's~t distribution for greater robustness to heavy tails, a direction applicable to our bivariate specification. \textcite{chen_generalization_2018} extend the Heckman selection model to a trivariate normal structure incorporating callback information, illustrating how additional survey process data could be accommodated within joint-normal latent-variable frameworks.}and parameters $(\alpha, \beta, \lambda, \theta,\rho)$ of model (\ref{eq:model}) are estimated via maximum likelihood, while parameters $\lambda_{Y-1}$ and $\theta_R$ are normalized to be equal to zero. The log-likelihood function takes the following form: \[ \begin{aligned} \log L(\alpha,\beta,\lambda,\theta,\rho \mid y_n,r_n,x_n,z_n) &= \sum_{n=1}^N \sum_{r=1}^R \sum_{y=1}^Y \mathbf{1}\{r_n = r,\, y_n = y\} \\ &\quad \times \log \int \mathbf{1}\{\lambda_{y-1} \leq \alpha^T x_n + \epsilon \leq \lambda_{y},\, \theta_{r-1} \leq \beta^T z_n + \eta \leq \theta_{r}\} \\ &\quad \quad \times \phi(\epsilon,\eta)\, d\epsilon\, d\eta \\[0.5em] &\quad + N_{\text{miss}} \cdot \log \sum_{k=1}^K p_k^z \int \mathbf{1}\{\beta^T z_k + \eta \geq \theta_R\}\, \phi(\eta)\, d\eta, \end{aligned} \] Standard errors are derived using the delta method with numerical Jacobian matrix. This method essentially uses a parametric model to extrapolate from low probability responders into nonresponders. While relying on a parametric model may seem restrictive, as \textcite{peress2010correcting} noted, {\it some} amount of extrapolation is necessary regardless of its technical form. Simulations in \textcite{peress2010correcting} documented meaningful gains in the case of a binary outcome. In Online Appendix~C, we show that the proposed estimator improves over a naïve ordinal probit model (which assumes MAR) and over the binarized version of the binary VRP approach of \textcite{peress2010correcting} across a range of scenarios.

\subsection{Identification and interpretation}
Identification of $\rho$ is driven by systematic variation in the
distribution of $y_n$ across categories of the proxy $r_n$ among
respondents; information about unit nonresponse (via $N_{\text{miss}}$
or a sensitivity grid) determines the implied mass of nonrespondents
when forming population quantities. The non-response rate set by the user corresponds to $N_{\text{miss}}/(N+N_{\text{miss}}).$  We illustrate the method under hypothetical nonresponse rates of 50\%, 65\%, and 80\%.

Following \textcite{peress2010correcting},
the ordered-probit threshold structure on both equations together with
the unit-variance normalization on the joint shocks pins down $\rho$
without requiring a variable in $z_n$ that is excluded from the outcome
equation; the baseline application accordingly sets $z_n=x_n$. When the
respondent fraction is small and latent selection is strong,
identification leans more on the parametric structure (Online
Appendix~C); in such regimes, including in $z_n$ a predictor that
affects response but is plausibly excluded from the outcome -- contact
difficulty, call attempts, interviewer-coded cooperativeness, or
fieldwork-effort indicators -- can sharpen identification but is not
required. Monte Carlo simulations in Online Appendix~C confirm that the estimator recovers population shares and 
$\rho$ with low bias across a range of nonresponse rates and selection strengths.
Crucially, $\rho$ is estimated rather than assumed: it captures any tendency for respondents with worse outcomes to also be less (or more) willing to cooperate.

\section{Empirical illustration}

\begin{figure}[!htbp]
    \centering
    \includegraphics[height=0.4\textheight]{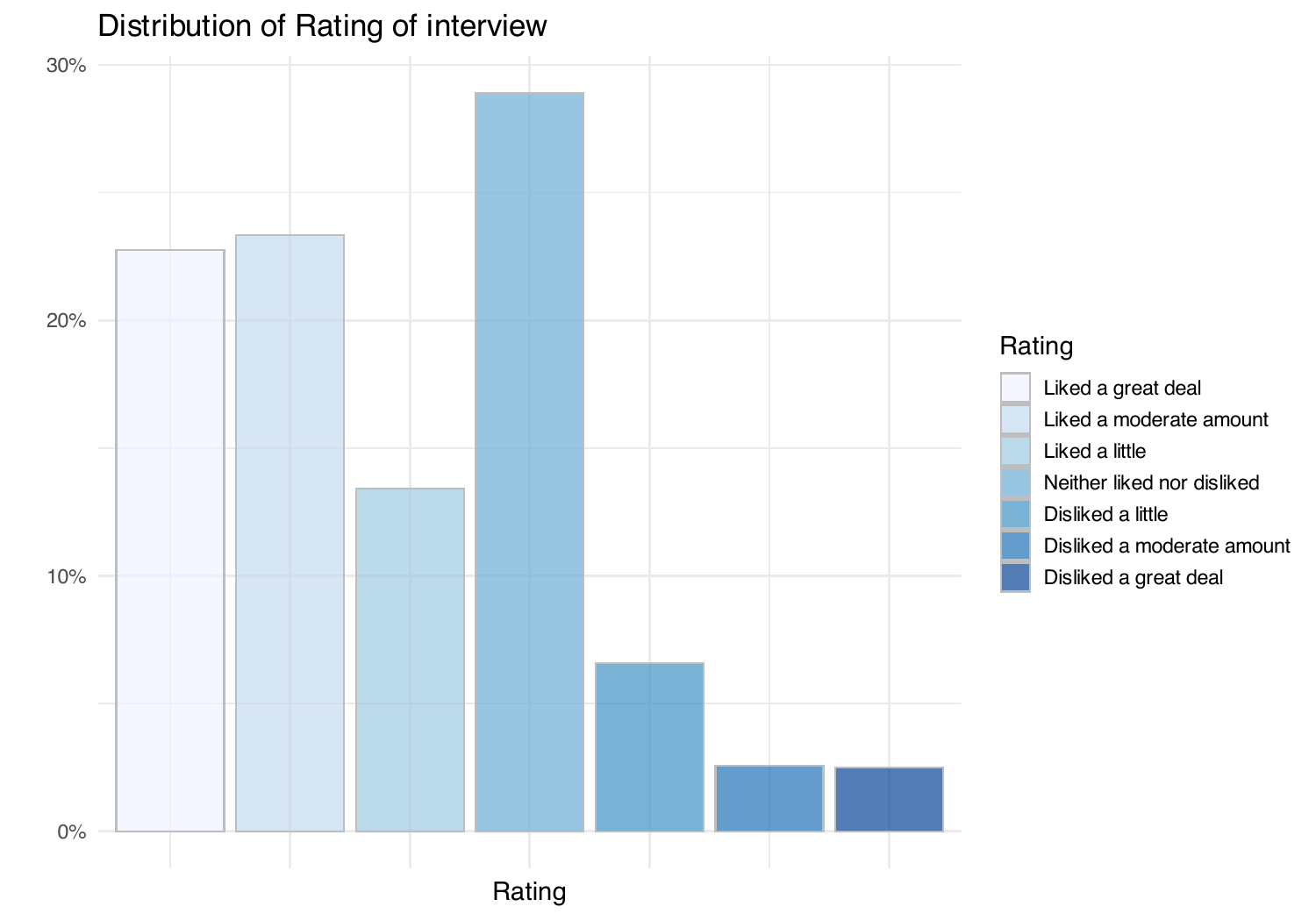}
    \caption{Distribution of the response variable {\it Rating of interview}.}
    \label{fig:response}
\end{figure}

We illustrate the proposed method using the Preliminary Release of the 2024 American National Election Studies (dated February 19, 2025) \citep{ANES2025}. 
The dataset contains approximately 3{,}000 respondents. The 2024 ANES Time Series Study reports AAPOR RR1 response rates of  33.4\% (in-person) and 37.7\% (web) \citep{ANES2025}, 
corresponding to non-response rates of roughly 67\% and 62\%. We therefore use 65\% nonresponse rate for the headline results.
As a response-propensity proxy we use the respondent's own {\it rating of the interview} with seven levels and Figure \ref{fig:response} captures its distribution. Other potential candidates for response variables such as {\it rating of the interviewer} or {\it do you take survey seriously} were not used due to their limited variability (see Online Appendix~A). The covariates consist of marital status, spouse’s gender (three categories: male, female, or inapplicable), race, and education (five levels), which together define 60 strata.

Following \textcite{peress2010correcting}, we set $z_n=x_n$ in the
main specification. This preserves the comparison with the original
binary VRP estimator and avoids restricting the analysis to subsamples
defined by mode-specific paradata; Online Appendix~D reports a
robustness check that adds a candidate exclusion-restriction variable
to $z_n$ and shows the conclusions are unchanged.

\begin{figure}[H]
    \centering
    {\large\bfseries How satisfied are you with life?}\\[6pt]
    \includegraphics[height=0.3\textheight,width=\linewidth,
  keepaspectratio]{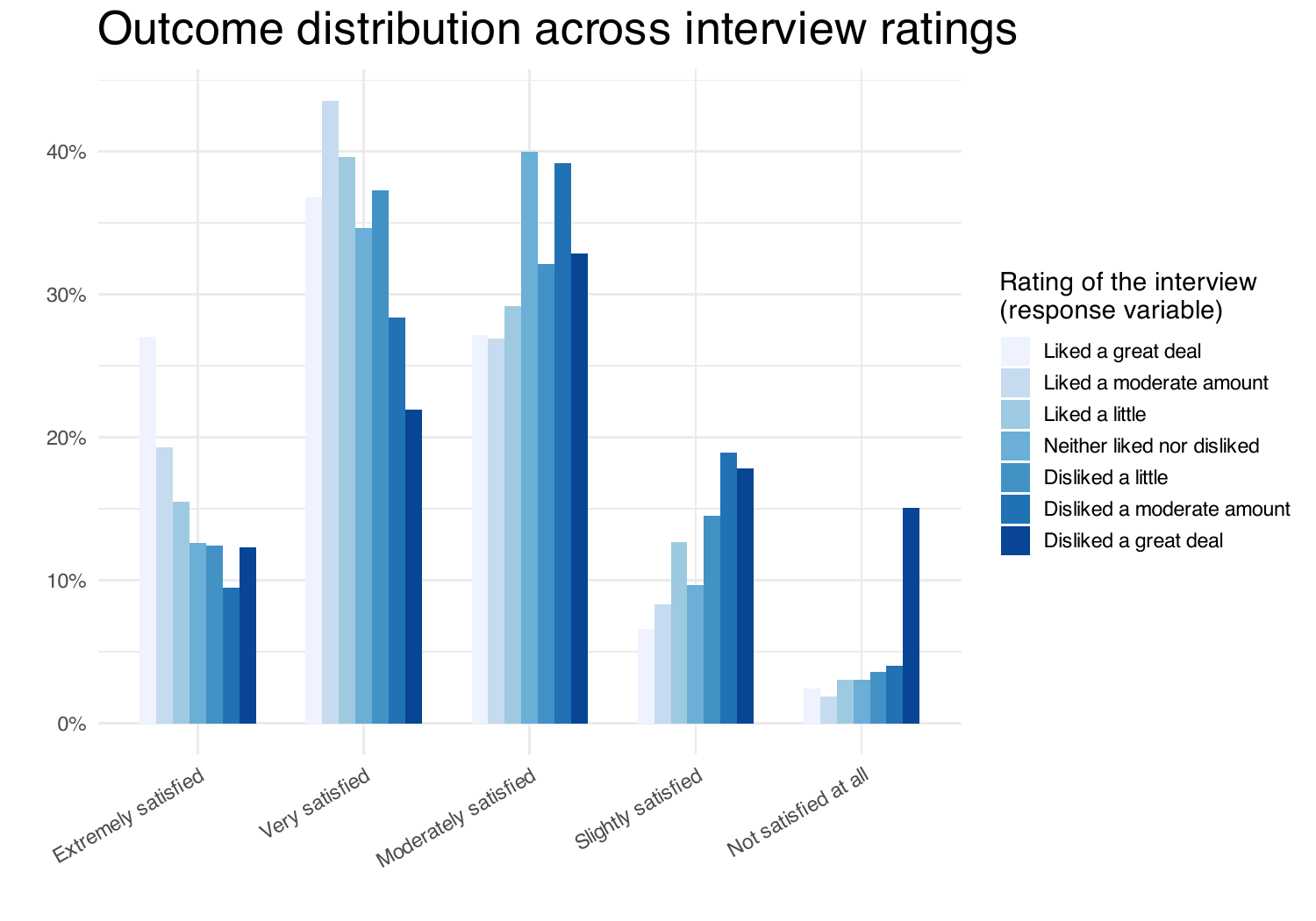}
    \includegraphics[height=0.3\textheight]{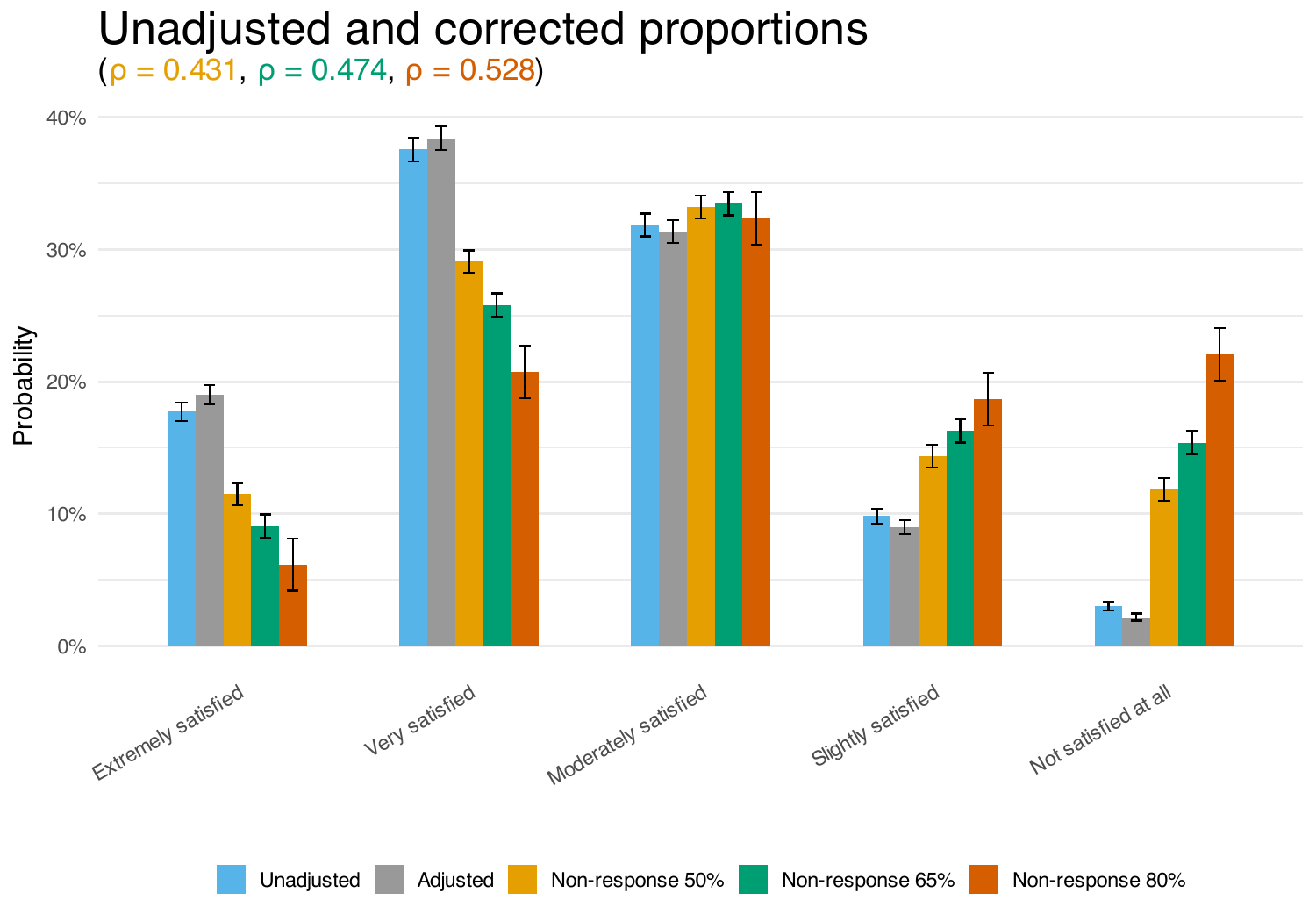}
    \includegraphics[height=0.3\textheight]{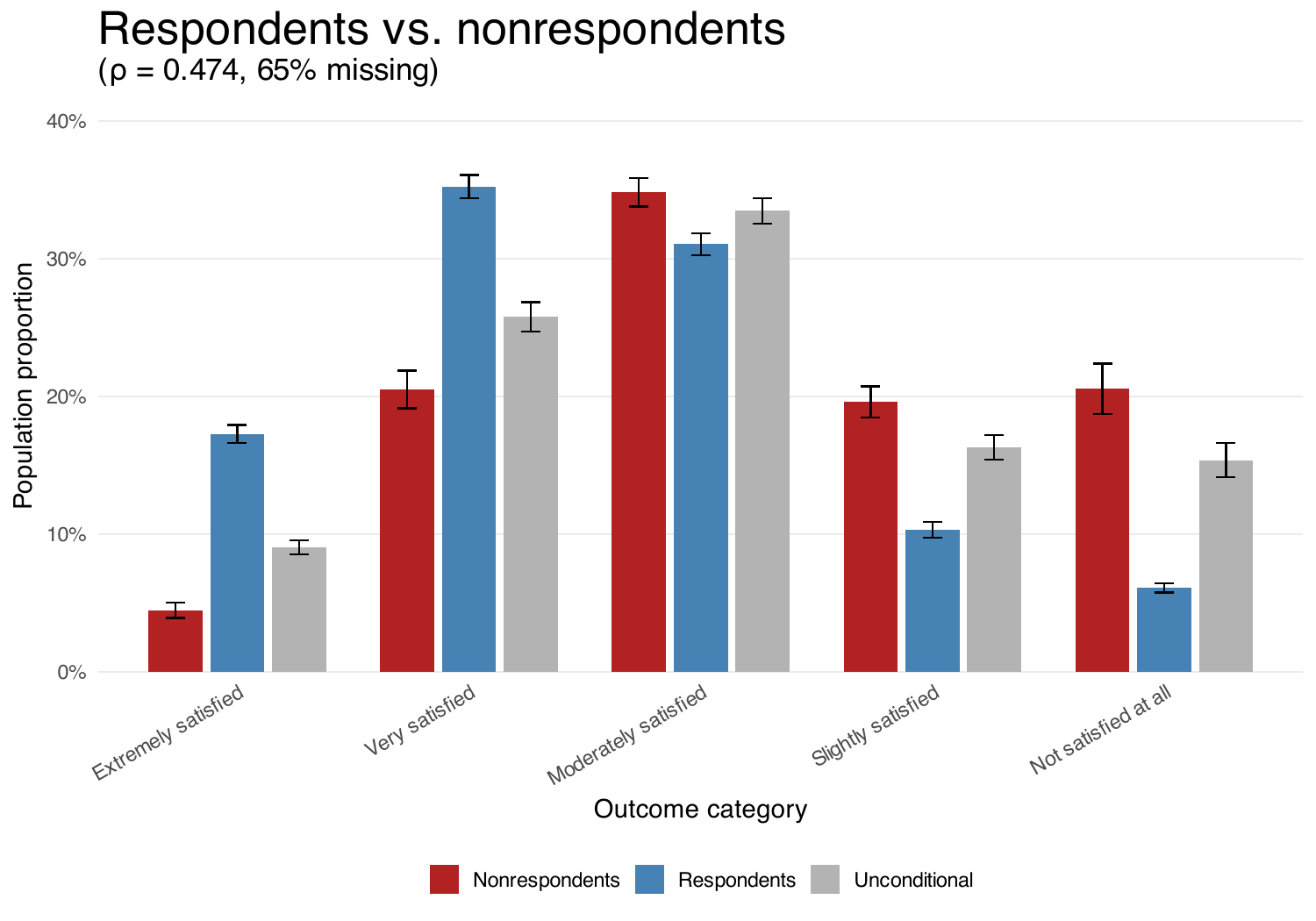}
    \caption{Question {\it How satisfied are you with life?} The upper panel shows the distribution of the response to this question across different ratings of the interview. Middle pane shows the unadjusted raw proportions, survey-weighted proportions and proportions estimated with the method proposed in this paper for different nonresponse rates: 50\%, 65\%, 80\%. Lower pane shows the comparison of the estimated distributions for the overall population, nonrespondents and respondents with 65\% non-response rate.}
    \label{fig:life}
\end{figure}

\begin{figure}[H]
    \centering
    {\large\bfseries Has the national economy gotten better or worse?}\\[6pt]
    \includegraphics[height=0.3\textheight]{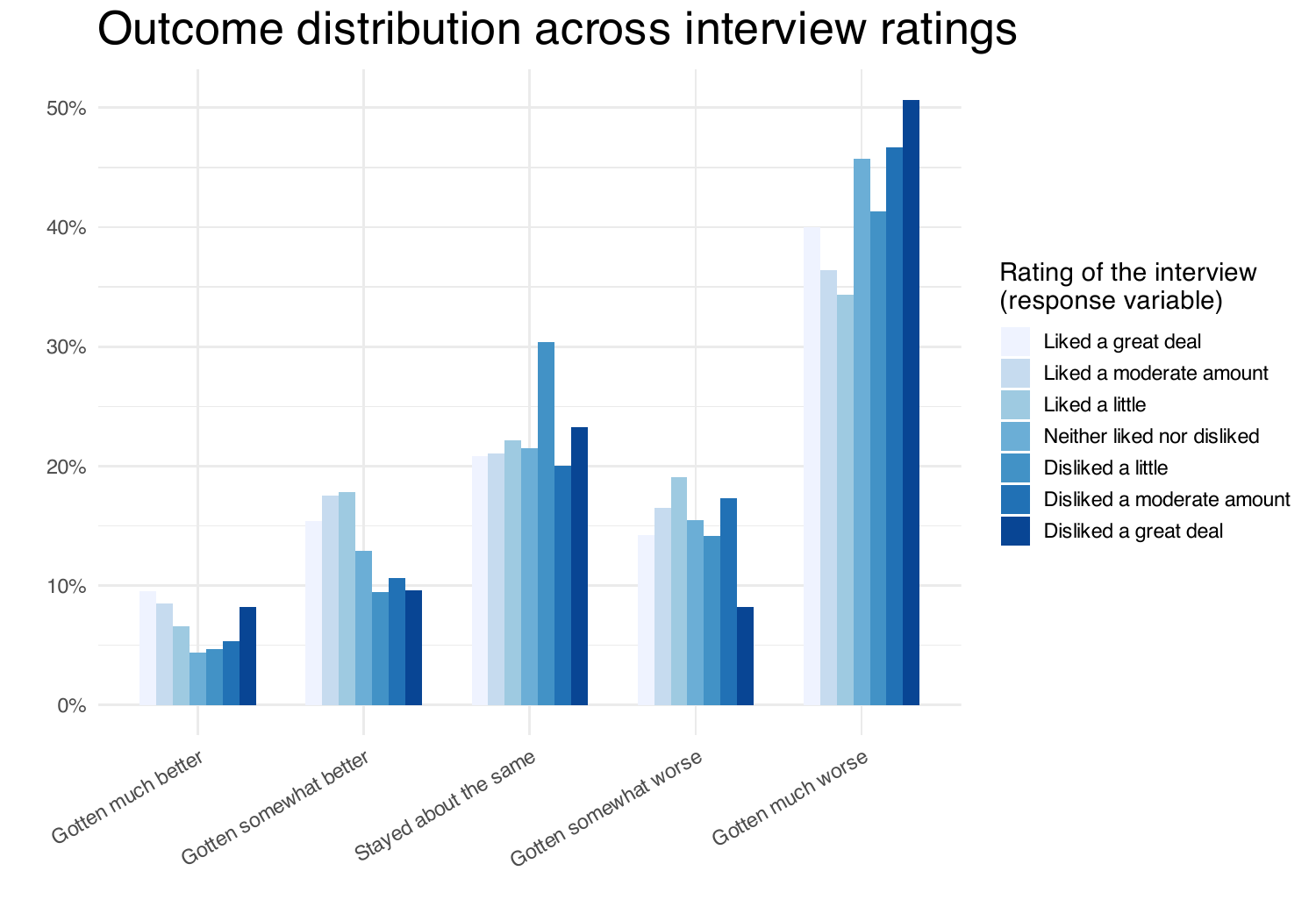}
    \includegraphics[height=0.3\textheight]{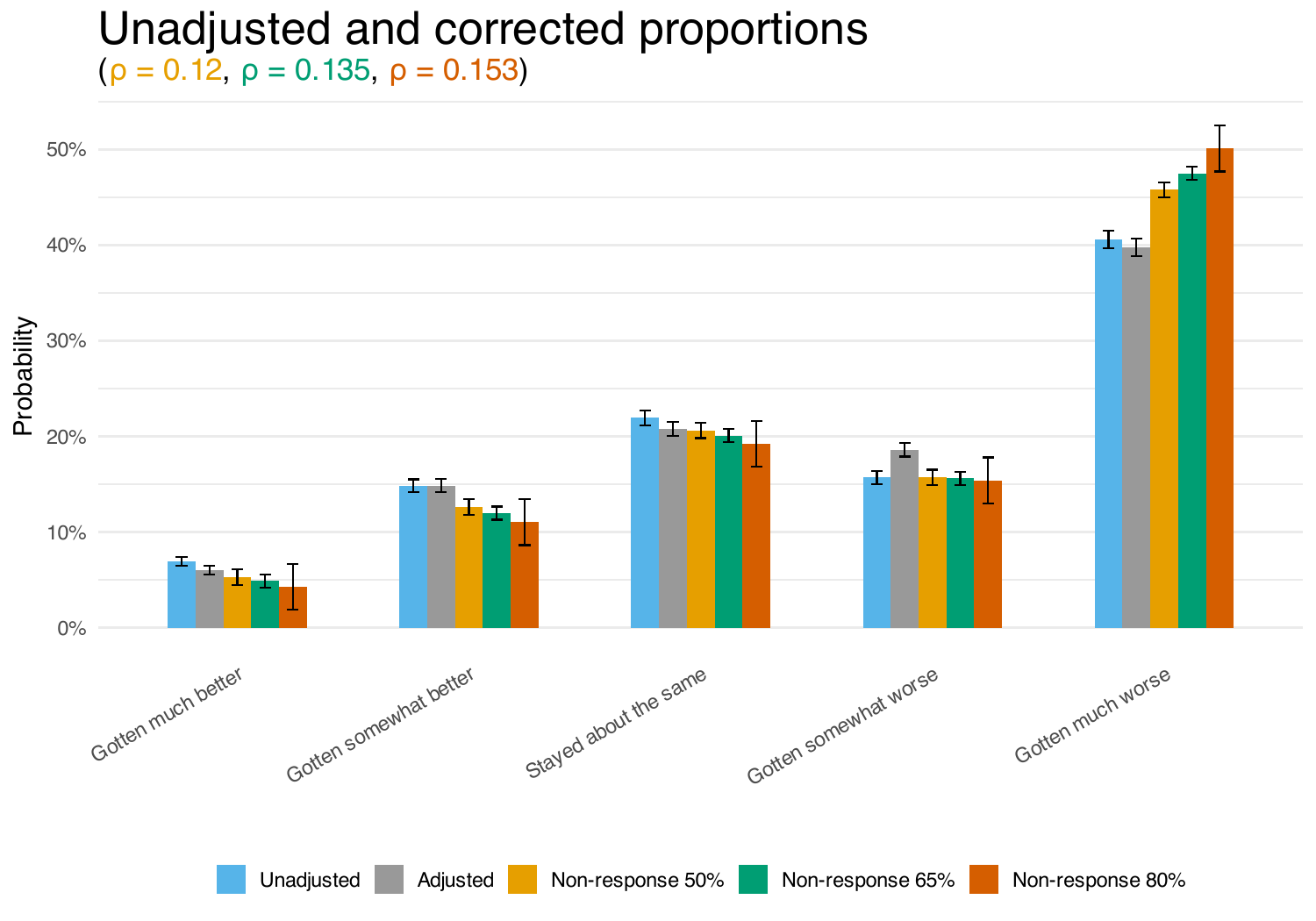}
    \includegraphics[height=0.3\textheight]{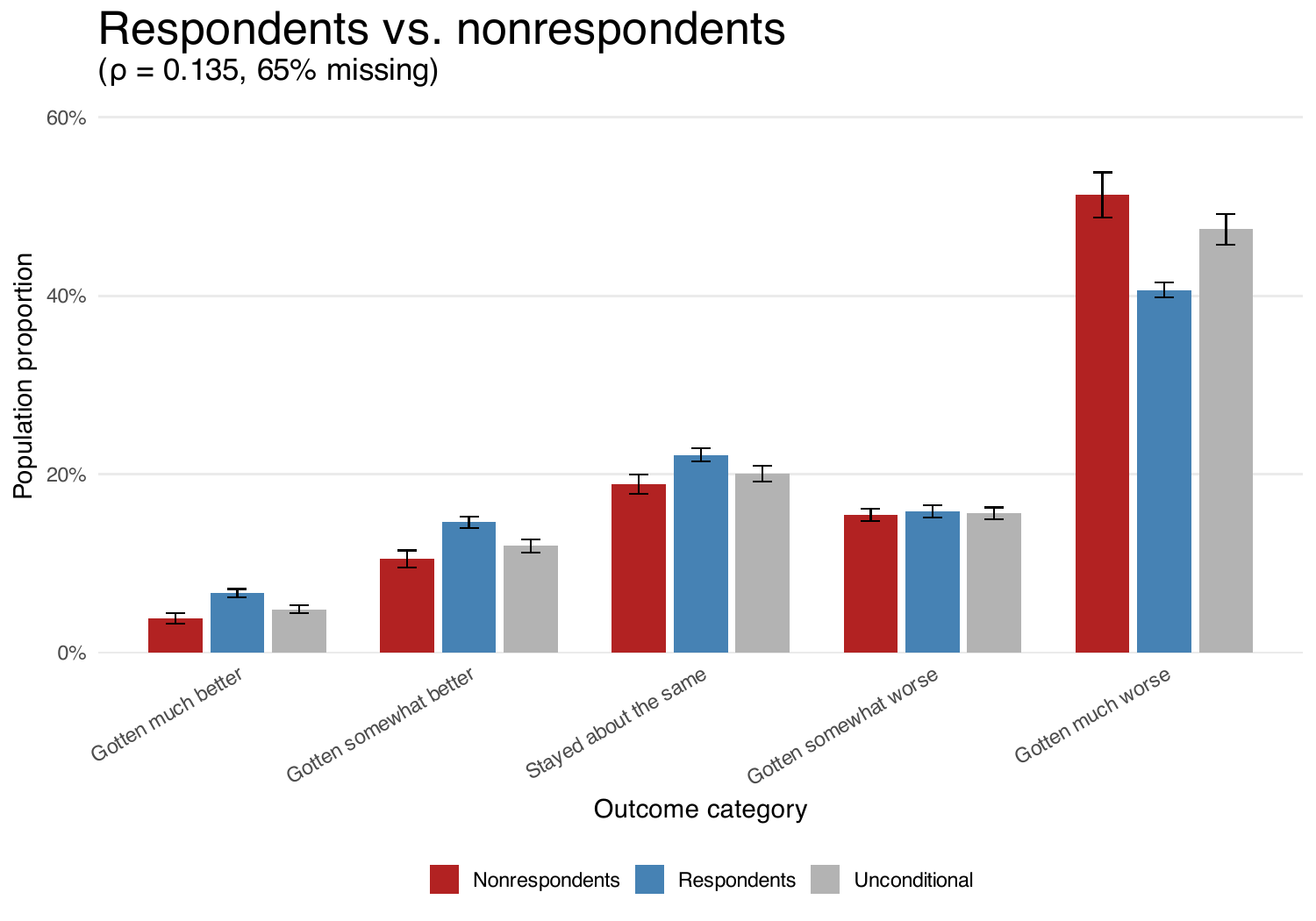}
    \caption{Question {\it Has national economy gotten better or worse?} 
    The upper panel shows the distribution of the response to this question across different ratings of the interview. 
    The middle panel shows the unadjusted raw proportions, survey-weighted proportions, and proportions estimated with the method proposed in this paper for different nonresponse rates: 50\%, 65\%, 80\%. 
    The lower panel shows the comparison of the estimated distributions for the overall population, nonrespondents, and respondents with a 65\% non-response rate.}
    \label{fig:economy}
\end{figure}

We use the respondent's self-reported rating of the interview as the VRP proxy for response propensity. In the same spirit, \textcite{peress2010correcting} uses an interviewer-coded measure of cooperativeness for this role; a respondent's own rating of the interview plays the same part, ordering respondents by willingness to cooperate and providing leverage for extrapolating from low-propensity respondents toward nonrespondents. In our application, the proxy displays substantial dispersion (Figure~\ref{fig:response}), and we assess the implied monotonic relationship by examining outcome distributions across proxy categories (upper panels of Figures~\ref{fig:life}--\ref{fig:economy}). 

Here we present two different survey questions from ANES, the relationship between the outcome and response measures and then distribution of the responses with varying degree of nonresponse together with estimated correlation coefficients. 

\begin{itemize}
\item Question 1: "How satisfied are you with life?"
\item Question 2: "Has national economy gotten better or worse?"
\end{itemize}

In the first question, we observe that respondents not satisfied with their life disliked the interview more (upper panel of Figure~\ref{fig:life}). This is reflected in the fact that estimates with high nonresponse  assign a much larger proportion to the {\it Slightly satisfied} or {\it Not satisfied at all} categories with estimated correlation coefficients around $0.47$. We see a mirror image on the other side of the distribution of life satisfaction. Also, we note that the proportion of respondents in the {\it Moderately satisfied} is barely affected. The model also allows us to estimate the outcome distribution separately for respondents and nonrespondents by conditioning on the latent response propensity (see Online Appendix~B). We observe a large difference between the nonrespondents and respondents. In this scenario, adjustment plays an important role. The life-satisfaction correction illustrates a pattern flagged by prior work on nonresponse in subjective well-being surveys: \textcite{heffetz_conclusions_2013} show that conclusions drawn from well-being data are sensitive to the difficulty of reaching respondents, and \textcite{hudomiet_age_2021} document that differential nonresponse biases the age profile of life satisfaction upward in the Health and Retirement Study. We present this result as an illustration of sensitivity to nonignorable nonresponse rather than a definitive reassessment, and further investigation is warranted.

Figure~\ref{fig:economy} shows the results for the question on perception of the national economy. Despite variability in the response-outcome relationship (upper panel) the nonresponse–adjusted proportions shift only modestly toward the {\it gotten worse} end, with an estimated correlation around $0.14$. The adjustment is far smaller than for life satisfaction, illustrating that the magnitude of the correction is outcome-specific.

Table~\ref{tab:summary} summarizes the adjustment across all eight
outcomes, adding questions on unemployment, media, elections, religion,
abortion, and the death penalty to the two discussed above; these are
illustrated graphically in Online Appendix~A. The estimated selection
correlation $\widehat\rho$ ranges from
essentially zero to about $0.5$. It is largest for life satisfaction
($0.47$) and the importance of religion ($0.26$): for these outcomes the
corrected distribution shifts toward less satisfaction with life and
lower religious importance, reflecting that more reluctant respondents,
and by extrapolation nonrespondents, differ systematically from eager
respondents. The adjustment is moderate for unemployment ($0.19$) and
the national economy ($0.14$), and statistically indistinguishable from
zero for media trust, vote-count accuracy, abortion importance, and the
death penalty, where the corrected and survey-weighted distributions
nearly coincide. The sign of $\widehat\rho$ records the direction of
selection: a positive value indicates that respondents who were more
reluctant tend toward higher outcome categories. These patterns mirror
the outcome-by-proxy relationships in Online Appendix~A and make clear
that the value of correcting for nonignorable nonresponse is
outcome-specific: substantial for a few attitudinal measures and
negligible for most.

\newpage

\begin{table}[h]
\centering
\footnotesize
\setlength{\tabcolsep}{5pt}
\caption{Summary of survey-weighted unadjusted and VRP-corrected population shares
(percentage points, rounded) for all eight outcomes, assuming
65\% nonresponse. $\hat\rho$
estimates the latent selection correlation; 95\% confidence intervals
in brackets.}
\label{tab:summary}
\begin{tabular}{llccrc}
\toprule
Outcome & Scale & Unadjusted & VRP-corrected & $\hat\rho$ & 95\% CI \\
\midrule
Life satisfaction
  & satisfied $\leftrightarrow$ not satisfied
  & 19 / 38 / 31 / 9 / 2 & 9 / 26 / 33 / 16 / 15
  & $\phantom{-}0.47$ & $[0.42,\;0.53]$ \\
National economy
  & got better $\leftrightarrow$ got worse
  & 6 / 15 / 21 / 19 / 40 & 5 / 12 / 20 / 16 / 47
  & $\phantom{-}0.14$ & $[0.06,\;0.21]$ \\
Unemployment
  & now better $\leftrightarrow$ now worse
  & 9 / 18 / 44 / 17 / 13 & 7 / 15 / 40 / 17 / 22
  & $\phantom{-}0.19$ & $[0.11,\;0.26]$ \\
Media trust
  & none $\leftrightarrow$ a great deal
  & 21 / 33 / 32 / 11 / 4 & 21 / 27 / 31 / 11 / 10
  & $\phantom{-}0.05$ & $[-0.02,\;0.12]$ \\
Votes counted accurately
  & not at all $\leftrightarrow$ completely
  & 9 / 11 / 29 / 34 / 17 & 12 / 10 / 28 / 31 / 19
  & $-0.05$ & $[-0.12,\;0.02]$ \\
Religion important
  & extremely $\leftrightarrow$ not at all
  & 28 / 19 / 20 / 14 / 20 & 21 / 16 / 19 / 15 / 30
  & $\phantom{-}0.26$ & $[0.19,\;0.33]$ \\
Abortion important
  & not at all $\leftrightarrow$ extremely
  & 4 / 9 / 25 / 30 / 31 & 4 / 10 / 26 / 28 / 32
  & $-0.03$ & $[-0.10,\;0.05]$ \\
Death penalty
  & favor $\leftrightarrow$ oppose
  & 39 / 22 / 19 / 20 & 39 / 21 / 19 / 20
  & $\phantom{-}0.03$ & $[-0.05,\;0.11]$ \\
\bottomrule
\end{tabular}
\end{table}
\section{Conclusion}
In this letter, we provide a practical method capable of correcting for nonignorable nonresponse bias, that generalizes the variable-response-propensity (VRP) framework of \textcite{peress2010correcting} from binary to also include ordinal outcomes. The proposed estimator combines (i) an ordinal response-propensity proxy observed among respondents, (ii) covariates defining post-stratification cells with known population shares, and (iii) a correlated latent-variable selection structure. The method is implemented in a compact \textsf{R} routine. Estimation is computationally light and can be run on a standard laptop, in our ANES illustration the full estimation run takes about ten minutes.

Two practical implications follow. First, the value of correcting for nonignorable nonresponse is outcome-specific and empirically assessable: in our illustration, adjustment meaningfully changes the distribution of life satisfaction but produces only a modest change for retrospective economic evaluations. Second, the approach is most informative when the response-propensity proxy exhibits meaningful dispersion and is plausibly monotone in willingness to cooperate. Identification relies on the parametric threshold structure and does not require an exclusion restriction, though including in the response equation a predictor that plausibly affects response propensity but not the outcome can provide additional leverage, particularly at extreme
nonresponse rates.

As with all selection-based corrections, conclusions depend on the assumed latent structure and on the credibility of any exclusion restrictions. Nonetheless, by making the extrapolation from respondents to nonrespondents explicit, transparent, and easy to implement, the ordinal VRP estimator provides a practical tool for sensitivity-aware descriptive inference with widely used ordinal survey measures.

\paragraph{Acknowledgments}
We thank the editor and three anonymous reviewers for their valuable comments and suggestions. We also thank Michael Peress for sharing his computer code with us, and Ilker Kalin and participants at the MIER 2025 and SEAM 2025 conferences for helpful comments and suggestions. We acknowledge the use of OpenAI's GPT-5 and Anthropic’s Claude for assistance with grammar, language and code editing. The content, analysis, and interpretations presented in the letter, online appendix, and replication files are entirely our own. Any remaining errors are ours.

\paragraph{Funding Statement}
This research was supported by Horizon Europe (award 101079219 and 101177204) and the Recovery and Resilience Plan for Slovakia (award 09I01-03-V04-00063/2024/VA). L.L. acknowledges support from the Slovak Research and Development Agency under contracts VEGA 
1/0645/26 and APVV-21-0360.

\paragraph{Competing Interests}
The authors declare no conflict of interest.

\paragraph{Data Availability Statement}
Replication data and code can be found at:\newline \url{https://github.com/LukasLaffers/vrpoprob}.

\paragraph{Author Contributions}
 Conceptualization: L.L.; J.M.M.; I.S. Methodology: L.L.; J.M.M.; I.S. Data curation: L.L.; J.M.M.; I.S. Writing original draft: L.L.; J.M.M.; I.S. All authors approved the final submitted draft.

\printbibliography

\includepdf[pages=-,fitpaper=true]{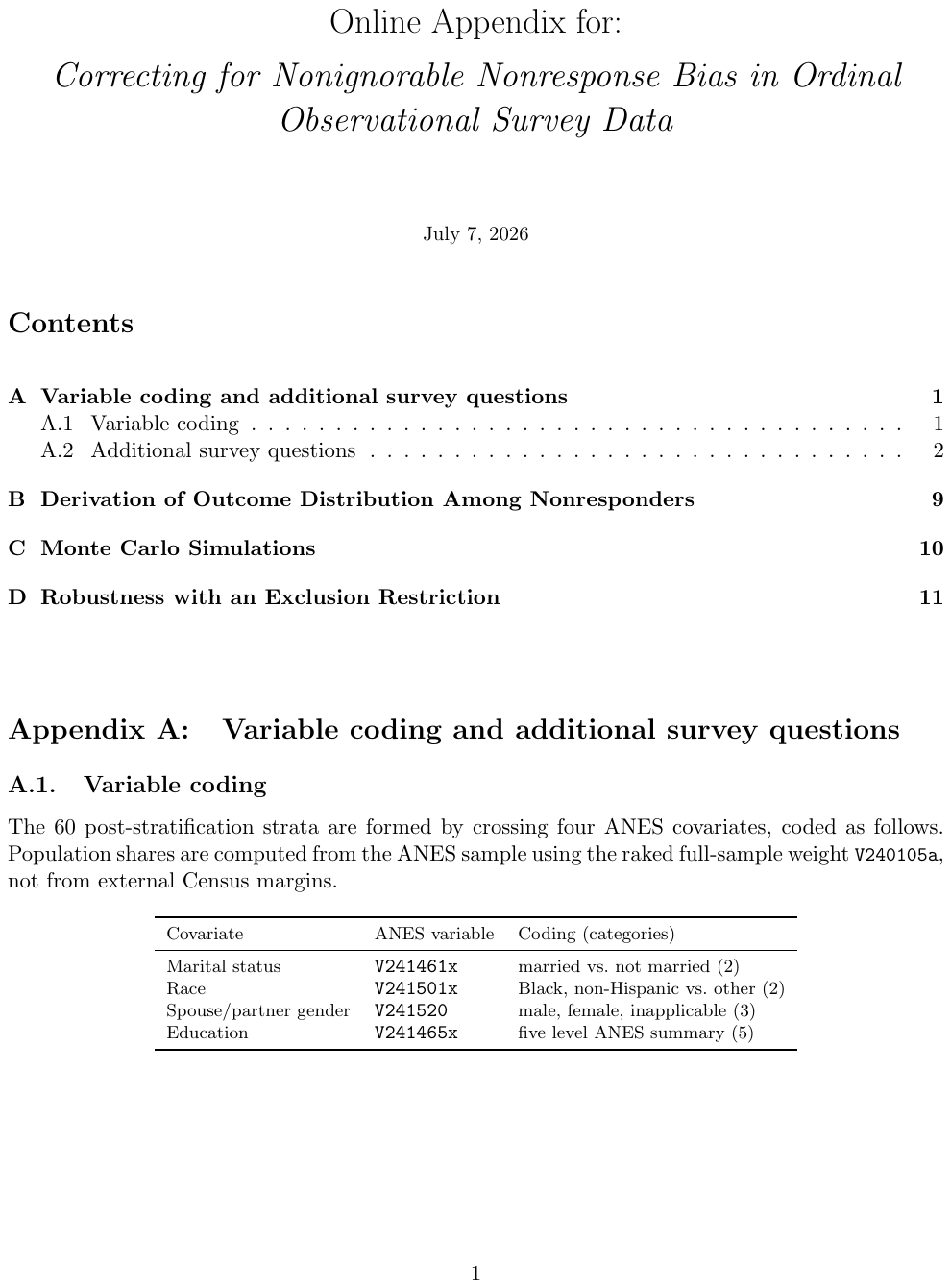}

\end{document}